\documentstyle[12pt]{article}

\begin{document}
\begin{flushright}
hep-th/0512334\\BHU/December/2005
\end{flushright}

\vspace{0.5cm}

\begin{center}

 {\bf \Large Augmented Superfield Approach To
 Exact Nilpotent Symmetries For Matter Fields In
 Non-Abelian Theory}

\vspace{1.5cm}

R. P. Malik \footnote{On leave of absence from S. N. Bose National
Centre for Basic Sciences, Salt Lake, Kolkata and presently
associated with the Centre of Advanced Studies, Department of
Physics, BHU, Varanasi. E-mail addresses: malik@bhu.ac.in ;
malik@bose.res.in} and Bhabani Prasad Mandal\footnote{E-mail
addresses: bhabani@bhu.ac.in ; bpm@bose.res.in}

Department of Physics, Banaras Hindu University, Varanasi-221005,
India

\vspace{0.7cm}

{\bf Abstract}

\end{center}

We derive the nilpotent (anti-) BRST symmetry transformations for
the Dirac (matter) fields of an interacting four
$(3+1)$-dimensional 1-form non-Abelian gauge theory by applying
the theoretical arsenal of augmented superfield formalism where
(i) the horizontality condition,  and (ii) the equality of a gauge
invariant quantity, on the six (4, 2)-dimensional supermanifold,
are exploited together. The above supermanifold is parameterized
by four bosonic spacetime coordinates $x^\mu$  (with $\mu =
0,1,2,3 )$ and a couple of Grassmannian variables $\theta $ and
$\overline{\theta}$. The on-shell nilpotent BRST symmetry
transformations for all the fields of the theory  are derived by
considering the chiral superfields on the five ($4,
1)$-dimensional super sub-manifold and the off-shell nilpotent
symmetry transformations emerge from the consideration of the
general superfields on the full six (4, 2)-dimensional
supermanifold. Geometrical interpretations for all the above
nilpotent symmetry transformations are also discussed in the
framework of augmented superfield formalism.

\section{Introduction}

The usual superfield approach [1-6] to Becchi-Rouet-Stora-Tyutin
(BRST) formalism provides the geometrical origin and
interpretations for some of the abstract mathematical properties
associated with the nilpotent and conserved (anti-) BRST charges
(and the nilpotent symmetry transformations they generate) for the
gauge and (anti-) ghost fields of the $p$-form ($p = 1, 2, 3...$)
interacting gauge theories. This approach, however, does not shed
any light on the (anti-) BRST symmetry transformations (and the
corresponding generators) for the matter fields that have
interactions with the $p$-form gauge fields of the above
interacting gauge theories. It has been an interesting and
challenging problem to derive these symmetry transformations for
the matter fields in the framework of superfield formalism without
spoiling the beauty of the geometrical interpretations (for the
conserved and nilpotent (anti-) BRST charges, the corresponding
nilpotent transformations for the gauge and (anti-) ghost fields,
their associated key properties, etc.) which emerge from the
horizontality condition {\it alone}.

To elaborate a bit on the usual superfield approach to BRST
formalism (endowed with the theoretical arsenal of horizontality
condition {\it alone}), it will be noted that, for the
$D$-dimensional Abelian $p$-form gauge theories, a $(p + 1)$-form
super curvature\footnote{ For the 1-form non-Abelian gauge theory,
the super curvature 2-form $\tilde F^{(2)} = \tilde d\tilde
A^{(1)} + i \tilde A^{(1)} \wedge A^{(1)}$ is equated with the
ordinary 2-form $F^{(2)} = d A^{(1)} + i A^{(1)} \wedge
A^{(1)}$due to the horizontality condition that leads to the
derivation of the nilpotent (anti-) BRST symmetry transformations
for the gauge and (anti-)ghost fields (see, e.g., [3] for
details).}
 $ \tilde F^{(p + 1)} = \tilde d \tilde
A^{(p)}$ is constructed from the super exterior derivative $\tilde
d = dx^\mu \partial_\mu + d \theta \partial_\theta + \bar\theta
\partial_{\bar\theta}$ (with $\tilde d^2 = 0$) and the super
$p$-form connection $\tilde A^{(p)}$ on a $(D, 2)$-dimensional
supermanifold parameterized by the $D$-number of bosonic spacetime
variables $x^\mu$ (with $\mu = 0, 1, 2....D-1)$ and a couple of
Grassmannian variables $\theta$ and $\bar \theta$ (with $\theta^2
= \bar\theta^2 = 0, \theta \bar\theta + \bar\theta \theta = 0$).
This is subsequently equated with the $D$-dimensional ordinary $(p
+ 1)$-form curvature $F^{(p + 1)} = d A^{(p)}$ constructed from
the ordinary exterior derivative $d = dx^\mu \partial_\mu$ (with
$d^2 = 0$) and the ordinary $p$-from connection $A^{(p)} =
\frac{1}{p!} (dx^{\mu_1} \wedge dx^{\mu_2}.......\wedge
dx^{\mu_p}) A_{\mu_1, \mu_2......\mu_p}$ due to the horizontality
condition (which has been christened as the soul-flatness
condition in [7]). This leads to the derivation of the nilpotent
and anticommuting (anti-) BRST symmetry transformations for {\it
only} the gauge and (anti-) ghost fields of the $D$-dimensional
Lagrangian density of an (anti-) BRST invariant $p$-form Abelian
gauge theory.

In a recent set of papers [8-13], the above horizontality
condition has been consistently extended\footnote{These extended
version have been christened as the augmented superfield
formulation.} by requiring the equality of (i) the conserved
currents/charges, and (ii) the gauge (i.e. BRST) invariant
quantities  that owe their origin to the (super) covariant
derivatives on the appropriate supermanifolds, so that the
nilpotent (anti-) BRST symmetry transformations for the matter (or
its analogue) fields could be derived in a logical fashion. The
former restriction leads to the {\it consistent} derivation of the
nilpotent symmetry transformations for the matter fields whereas
the latter restriction yields the mathematically {\it exact}
nilpotent symmetry transformations for the matter fields. Both
these extensions of the usual superfield formalism have their own
merits. For instance, the former is applicable to any
reparametrization and/or gauge invariant theories where the
covariant derivatives do not play so important role (e.g. the
system of (super) relativistic particles [12,13]). The latter
crucially depends on the existence of covariant derivatives in the
theory (e.g. the $U(1)$ and $SU(N)$ gauge invariant theories
[14,15,11]).

One of the key points of the above extensions is the fact that the
geometrical interpretations for the (anti-) BRST charges (and the
corresponding nilpotent symmetry transformations generated by
them) remain exactly the same as in the usual superfield approach
to BRST formalism [1-7]. For instance, the above (anti-) BRST
charges turn out to be the translational generators (i.e.
($\mbox{Lim}_{\bar\theta \to 0} (\partial/\partial \theta)$)
$\mbox{Lim}_{\theta \to 0} (\partial/\partial \bar\theta)$) along
the Grassmannian directions of the $(D, 2)$-dimensional
supermanifold. Their nilpotency property is found to be encoded in
the two successive translations (i.e $(\partial/\partial \theta)^2
= 0, (\partial/\partial \bar\theta)^2 = 0$)  of the superfields
along any particular Grassmannian direction of the suitably chosen
$(D,2)$-dimensional supermanifold. The anticommutativity property
of the BRST and anti-BRST charges are found be linked with such a
kind of property (i.e. $(\partial/\partial \theta)
(\partial/\partial \bar\theta) + (\partial/\partial \bar\theta)
(\partial/\partial\theta) = 0$) associated with the translational
generators along the Grassmannian directions. Finally, the
internal BRST and anti-BRST symmetry transformations for the local
$D$-dimensional {\it ordinary} fields are geometrically equivalent
to the translations of the corresponding {\it superfields} along
the Grassmannian directions of the $(D, 2)$-dimensional
supermanifold.

In our present investigation, we exploit the latter extension of
the usual superfield approach to BRST formalism to obtain the
mathematically exact nilpotent (anti-) BRST symmetry
transformations for the matter (Dirac) fields of the physical four
$(3 + 1)$-dimensional (4D) non-Abelian gauge theory. First, as a
warm-up exercise, we derive the on-shell nilpotent BRST symmetry
transformations for the matter as well as gauge and (anti-) ghost
fields by invoking the definition of the chiral superfields on the
five ($4, 1)$-dimensional chiral super sub-manifold of the general
$(4, 2)$-dimensional supermanifold. This exercise also provides,
in a subtle way, the logical reason behind the non-existence of
the on-shell nilpotent anti-BRST symmetry transformations for a
certain specific set of Lagrangian densities for the interacting
4D non-Abelian gauge theory. Later on, we derive the off-shell
nilpotent (anti-) BRST symmetry transformations for {\it all} the
fields of the 4D non-Abelian gauge theory by considering the
general superfields on the general six $(4, 2)$-dimensional
supermanifold. For this purpose, we tap the potential and power of
the horizontality condition as well as the additional
gauge-invariant restriction on the above supermanifold. We
demonstrate that the above restrictions are beautifully
intertwined and possess a common mathematical origin.

Our present endeavour is essential basically on three counts.
First, it has been a challenging problem to derive the nilpotent
(anti-) BRST symmetry transformations for the matter fields of any
arbitrary gauge theory in the framework of superfield approach to
BRST formalism without having any conflict with the key results
and observations of the usual superfield formalism [1-7]. Second,
to check the validity of the gauge-invariant restriction in
yielding mathematically exact nilpotent (anti-) BRST
transformations for the matter fields of an interacting 4D
non-Abelian $SU(N)$ gauge theory which was found to be true for
the interacting 4D Abelian $U(1)$ gauge theory \footnote{It is
self-evident that the latter gauge theory (i.e. Abelian) is the
limiting case of the former (i.e. non-Abelian) theory which is
theoretically more general in nature.} [14,15]. Finally, the ideas
proposed in our present investigation might turn out to be useful
in the derivation of the nilpotent (anti-) BRST symmetry
transformations for the reparametrization invariant theories of
gravitation where the covariant derivatives play very important
roles in generating the interaction terms. It is well known that
some of the key features of non-Abelian gauge theories are very
closely connected to a few central ideas in the theory of
gravitation (see, e.g., [16] for details).

The contents of our present paper are organized as follows. To set
up the notations and conventions, we discuss in Sec. 2, the bare
essentials of the (anti-) BRST symmetry transformations for the
four (3+1)-dimensional interacting non-Abelian gauge theory.
Section 3 is devoted to the derivation of the on-shell nilpotent
BRST symmetry transformations in the augmented superfield
formulation where only the chiral superfields are considered on
the five (4, 1)-dimensional super sub-manifold of the general six
(4, 2)-dimensional supermanifold. The off shell nilpotent (anti-)
BRST symmetry transformations for all the fields of the Lagrangian
densities  are derived in Sec. 4 where a general set of
superfields are considered on the six (4, 2)-dimensional
supermanifold. Finally, in Sec. 5, we make some concluding remarks
and point out a few future directions for further investigations.

\section{Nilpotent Symmetry Transformations in
Lagrangian formalism : A Brief Sketch }

We begin with the BRST invariant Lagrangian density of the
physical four $(3+1)$ dimensional non-Abelian 1-form (i.e.
$A^{(1)} = dx^\mu A_\mu$) interacting gauge theory, in the Feynman
gauge, as [16,17]
\begin{equation}
{\cal L }_b = - \frac{1}{4}F_{\mu\nu}\cdot F^{\mu\nu} +
\bar{\psi}(i\gamma^\mu D_\mu-m)\psi + B\cdot(\partial_\mu A^\mu) +
\frac{1}{2}B\cdot B - i\partial_\mu\bar{C}\cdot D^\mu
C,\label{2.1}
\end{equation}
where $ F_{\mu\nu}= \partial_\mu A_\nu -\partial_\nu A_\mu + i
A_\mu \times A_\nu $ is the field strength tensor for the group
valued non-Abelian gauge potential $A_\mu\equiv A_\mu ^ a T^a$
with the group generators $T$'s obeying the algebra $[T^a, T^b] =
f^{abc} T^c$. The structure constant $f^{abc}$ can be chosen to be
totally antisymmetric in the group indices $a, b $ and $c$ for a
semi simple Lie group [16]. The covariant derivatives $D_\mu \psi
= \partial_\mu \psi + i A_\mu ^a T^a \psi $ and $D_\mu C^a =
\partial_\mu C^a + i f^{abc} A_\mu ^b C^c \equiv \partial_\mu C^a + i
(A_\mu\times C)^a$ are defined \footnote{ We adopt here the
conventions and notations such that the Minkowski 4D flat metric
is $\eta_{\mu\nu}=$ diag $(+1, -1, -1, -1)$ on the spacetime
manifold. The dot product $A \cdot B= A^a B^a$ and cross product
$(A\times B )^a= f^{abc} A^b B^c$ are defined in the group space
of the semi simple Lie group. Here the Greek indices $\mu ,\nu ,
\rho...= 0, 1, 2, 3$ stand for the spacetime directions on the
ordinary Minkowski spacetime manifold and the Latin indices $a, b,
c ...=1, 2, 3 ...$ denote the $SU(N)$ group indices in the
``colour'' space of the internal group manifold.} on the matter
(quark) field $\psi $ and ghost field $C^a$ such that $[D_\mu ,
D_\nu]\psi= i F_{\mu\nu} \psi$ and $[D_\mu , D_\nu] C^a = i
(F_{\mu\nu} \times C)^a$. It will be noted that these definitions
for $F_{\mu\nu}$ agree with the Maurer-Cartan equation $F^{(2)} =
d A^{(1)}+i A^{(1)}\wedge A^{(1)}\equiv \frac{1}{2!} (dx^\mu
\wedge dx^\nu) F_{\mu\nu}$ that defines the 2-form $F^{(2)}$
which, ultimately, leads to the derivation of $F_{\mu\nu}$. In the
above equation (\ref{2.1}), $B^a$ are the Nakanishi-Lautrup
auxiliary fields and the anticommuting (i.e. $(C^a)^2 = 0 =
(\bar{C}^a)^2, C^a \bar{C}^b + \bar{C}^b C^a =0$) (anti-) ghost
fields $(\bar C^a) C^a$ are required for the proof of unitarity in
the 1-form non-Abelian gauge theory \footnote{ The importance of
the ghost fields appears in the proof of unitarity of a given
physical process (allowed by the interacting non-Abelian gauge
theory) where, for each gluon (bosonic non-Abelian gauge field)
loop (Feynman) diagram, a loop diagram constructed by the
fermionic (anticommuting) ghost field alone, is required (see,
e.g. [18] for details). }. Furthermore, the $\gamma$'s are the
usual $4 \times 4$ Dirac matrices in the physical four $(3 +
1)$-dimensional Minkowski space.

The above Lagrangian density (\ref{2.1}) respects the following
off-shell nilpotent ($s_b^2 = 0$) BRST symmetry transformations
($s_b$) [16,17]
\begin{eqnarray}
&&s_b A_\mu = D_\mu C, \quad s_b C = - \frac{i}{2} (C \times C),
\quad s_b \bar C = i B, \quad s_b B = 0, \nonumber\\ && s_b \psi =
- i (C \cdot T) \psi, \quad s_b \bar \psi = - i \bar \psi (C \cdot
T), \quad s_b F_{\mu\nu} = i (F_{\mu\nu} \times C).\label{2.2}
\end{eqnarray}
It will be noted that the kinetic energy term $(- \frac{1}{4}
F^{\mu\nu} \cdot F_{\mu\nu})$ remains invariant under the BRST
transformations (i.e. $s_b (-\frac{1}{4} F^{\mu\nu} \cdot
F_{\mu\nu}) = - \frac{i}{2} f^{abc} F_{\mu\nu}^a F^{\mu\nu b} C^c
= 0$) because of the totally antisymmetric property of the
structure constants $f^{abc}$. The on-shell ($\partial_\mu D^\mu C
= 0$) nilpotent ($\tilde s_b^2 = 0$) version of the above
nilpotent symmetry transformations (i.e. $\tilde s_b$), are
\begin{eqnarray}
&& \tilde s_b A_\mu = D_\mu C, \quad \tilde s_b C = - \frac{i}{2}
(C \times C), \quad \tilde s_b \bar C = - i (\partial_\mu A^\mu),
\nonumber\\ && \tilde s_b \psi = - i (C \cdot T) \psi, \quad
\tilde s_b \bar \psi = - i \bar \psi (C \cdot T), \quad \tilde s_b
F_{\mu\nu} = i (F_{\mu\nu} \times C),\label{2.3}
\end{eqnarray}
under which the following Lagrangian density
\begin{eqnarray}
&& \displaystyle {{\cal L}_{b}^{(0)} = - \frac{1}{4} F^{\mu\nu}
\cdot F_{\mu\nu} + \bar \psi \;(i \gamma^\mu D_\mu - m)\; \psi}
\nonumber\\ &&- \displaystyle {\frac{1}{2} (\partial_\mu A^\mu)
\cdot (\partial_\rho A^\rho) - i
\partial_\mu \bar C \cdot D^\mu C},\label{2.4}
\end{eqnarray}
changes to a total derivative (i.e. $\tilde s_b {\cal L}_b^{(0)} =
- \partial_\mu [ (\partial_\rho A^\rho) \cdot D^\mu C ])$.

The following off-shell nilpotent ($s_{ab}^2 = 0$) version of the
anti-BRST ($s_{ab}$) transformations
\begin{eqnarray}
&& s_{ab} A_\mu = D_\mu \bar C, \qquad s_{ab} \bar C = -
\frac{i}{2} (\bar C \times \bar C), \qquad s_{ab} C = i \bar B,
\nonumber\\&& s_{ab} B = i (B \times \bar C), \qquad s_{ab}
F_{\mu\nu} = i (F_{\mu\nu} \times \bar C),  \qquad s_{ab} \bar B =
0, \nonumber\\ && s_{ab} \psi = - i (\bar C \cdot T) \psi, \qquad
s_{ab} \bar \psi = - i \bar \psi (\bar C \cdot T),\label{2.5}
\end{eqnarray}
are the symmetry transformations for the following equivalent
Lagrangians
\begin{eqnarray}
{\cal L}_{\bar B}^{(1)} &=& - \displaystyle { \frac{1}{4}
F^{\mu\nu} \cdot F_{\mu\nu} + \bar \psi\; (i \gamma^\mu D_\mu - m
)\; \psi + B \cdot (\partial_\mu A^\mu)} \nonumber\\ &+&
\frac{1}{2} (B \cdot B + \bar B \cdot \bar B) - i \partial_\mu
\bar C \cdot D^\mu C,\label{2.6a}
\end{eqnarray}
\begin{eqnarray}
{\cal L}_{\bar B}^{(2)} &=& - \displaystyle { \frac{1}{4}
F^{\mu\nu} \cdot F_{\mu\nu} + \bar \psi \;(i \gamma^\mu D_\mu - m
)\; \psi - \bar B \cdot (\partial_\mu A^\mu)} \nonumber\\ &+&
\frac{1}{2} (B \cdot B + \bar B \cdot \bar B) - i D_\mu \bar C
\cdot \partial^\mu C,\label{2.6b}
\end{eqnarray}
where another auxiliary field $\bar B$ has been introduced with
the restriction $B + \bar B = - (C \times \bar C)$ (see, e.g.
[19]). It can be checked that the anticommutativity property ($s_b
s_{ab} + s_{ab} s_b = 0$) is true for any arbitrary field of the
above Lagrangian densities. For the proof of this statement, one
should also take into account $s_b \bar B = i (\bar B \times C)$
which is not mentioned in (\ref{2.2}). We emphasize that the
on-shell version of the anti-BRST symmetry transformations does
not exist for any of the above cited Lagrangian densities (see,
e.g. [20]).

All the above nilpotent symmetry transformations can be succinctly
expressed in terms of the conserved and off-shell nilpotent
(anti-) BRST charges $Q_r$ (with $r = b, ab$) and on-shell
nilpotent BRST charge $\tilde Q_b$, as
\begin{equation}
s_r \Phi = - i\; [\; \Phi, Q_r\; ]_{\pm}, \qquad r = b, ab, \qquad
\tilde s_b \tilde \Phi = - i\; [\; \tilde \Phi, \tilde Q_b \;
]_{\pm},\label{2.7}
\end{equation}
where the $(+)-$ signs, as the subscripts, on the square brackets
stand for the (anti-) commutator for the generic field $\Phi =
A_\mu, C, \bar C, \psi, \bar\psi, B, \bar B$ and $\tilde \Phi =
A_\mu, C, \bar C, \psi, \bar\psi$ (of the above appropriate
Lagrangian densities) being (fermionic) bosonic in nature. For our
discussions, the explicit forms of $Q_r$ (r = b, ab) and $\tilde
Q_b$  are not essential but these can be found in [16,17].

\section{On-shell Nilpotent BRST Symmetries:  Augmented
Superfield Approach}

In this section, first of all, we take the chiral superfields,
defined on the five $(4, 1)$-dimensional super submanifold of the
general six (4, 2)-dimensional supermanifold and derive the
on-shell nilpotent BRST symmetry transformations for the gauge-
and the (anti-) ghost fields by exploiting the well-known
horizontality condition. Later on, in Subsec. 3.2, we derive the
nilpotent BRST transformations for the Dirac  fields $\psi$ and
$\bar\psi$ by exploiting a gauge-invariant condition on the five
$(4, 1)$-dimensional chiral supermanifold.

\subsection{Horizontality Condition: On-shell Nilpotent
BRST Symmetries for Gauge and (Anti-) Ghost Fields}

For the present paper to be self-contained, we shall briefly
invoke the essentials of horizontality condition (connected with
the usual superfield formalism [1-7]) in obtaining the on-shell
nilpotent symmetry transformations (\ref{2.3}) that exist for the
non-Abelian gauge theory. For this purpose, we define the chiral
(i.e. $ \theta \to 0$) super exterior derivative and super 1-form
connection, as  (see, e.g. [20] for details):
\begin{equation}
\tilde d_{(c)} = dx^\mu\; \partial_\mu + d \bar\theta\;
\partial_{\bar\theta}, \qquad \tilde A^{(1)}_{(c)} = dx^\mu\;
B_\mu^{(c)} (x, \bar\theta) + d \bar\theta\; F^{(c)}
(x,\bar\theta). \label{3.1}
\end{equation}
The chiral super expansions for the following multiplet
superfields are [20]
\begin{eqnarray}
&& (B_\mu^{(c)}\cdot T) (x, \bar\theta) = (A_\mu \cdot T) (x) +
\bar \theta\; (R_\mu \cdot T) (x),\nonumber\\ && (F^{(c)} \cdot T)
(x, \bar\theta) = (C \cdot T) (x) + i\; \bar\theta \;(B_1 \cdot T)
(x), \nonumber\\ && (\bar F^{(c)}\cdot T) (x, \bar\theta) = (\bar
C \cdot T) (x) + i\; \bar\theta\; (B_2 \cdot T) (x).\label{3.2}
\end{eqnarray}
In fact, these superfields are the chiral limits of the general
super expansion that would be discussed in Subsec. 4.1. In
general, the super exterior derivative $\tilde d = dx^\mu
\partial_\mu + d \theta \partial_\theta + d \bar\theta
\partial_{\bar\theta}$ and super 1-form connection $\tilde A^{(1)}
= d Z^M A_M = dx^\mu B_\mu (x, \theta, \bar\theta) + d \theta \bar
F (x, \theta, \bar\theta) + d \bar\theta F (x, \theta,
\bar\theta)$. It is evident that, in chiral limit $( \theta \to
0$), we have the definition of the super exterior derivative and
super 1-form connection as given in (\ref{3.1}). In the chiral
expansion (\ref{3.2}), $R_\mu \cdot T \equiv R^a_\mu T^a, B_1
\cdot T \equiv B^a_1 T^a$ and $B_2 \cdot T \equiv B^a_2 T^a$ are
the group valued secondary fields. In fact, in the horizontality
condition (i.e. $\tilde F^{(2)}_{(c)} = F^{(2)}$), we shall equate
the super chiral 2-form $\tilde F^{(2)}_{(c)} = \tilde d_{(c)}
\tilde A^{(1)}_{(c)} + i \tilde A^{(1)}_{(c)} \wedge \tilde
A^{(1)}_{(c)}$, defined on the five (4, 1)-dimensional super
sub-manifold,  where
\begin{eqnarray}
\tilde d_{(c)} \tilde A^{(1)}_{(c)} &=& (dx^\mu \wedge dx^\nu)
(\partial_\mu B_\nu^{(c)}) + (dx^\mu \wedge d\bar\theta) [
\partial_\mu F^{(c)}- \partial_{\bar\theta} B_\mu^{(c)}] \nonumber\\
&-& (d\bar\theta \wedge d\bar\theta) (\partial_{\bar\theta}
F^{(c)}),\label{3.3}
\end{eqnarray}
\begin{eqnarray}
i \tilde A^{(1)} \wedge \tilde A^{(1)} &=& i\; (dx^\mu \wedge
dx^\nu) (B_\mu^{(c)} B_\nu^{(c)}) + i\; (dx^\mu \wedge d
\bar\theta) [B_\mu^{(c)}, F^{(c)}] \nonumber\\ &-& i\; (d
\bar\theta \wedge d \bar\theta) (F^{(c)} F^{(c)}),\label{3.4}
\end{eqnarray}
with the ordinary 2-form $F^{(2)} = d A^{(1)} + i A^{(1)} \wedge
A^{(1)} \equiv \frac{1}{2!} (dx^\mu \wedge dx^\nu) (F_{\mu\nu}
\cdot T)$ curvature defined on the ordinary 4D spacetime manifold.

The consequences of the above horizontality condition $\tilde
F^{(2)}_{(c)} = F^{(2)}$ on the  five $(4, 1)$-dimensional chiral
super sub-manifold, are
\begin{equation}
R_\mu = D_\mu C, \qquad B_1 = - \frac{1}{2}\; (C \times C), \qquad
B_1 \times C = 0.\label{3.5}
\end{equation}
The insertions of the above values into the expansion (\ref{3.2}),
{\it vis-{\`a}-vis} the on-shell nilpotent transformations
(\ref{2.3}), lead to
\begin{equation}
B_\mu^{(c)} (x, \bar\theta) = A_\mu (x) + \bar \theta \;(\tilde
s_b A_\mu (x)), \quad F^{(c)} (x, \bar\theta) = C (x) + \bar\theta
\;(\tilde s_b C (x)).\label{3.6}
\end{equation}
It will be noted that the secondary field $(B_2 \cdot T)$ in the
expansion of $\bar F^{(c)}$ is {\it not} determined by the above
horizontality condition. The equation of motion $ B +
(\partial_\mu A^\mu) = 0$, however, comes to our rescue if we
identify the secondary field $B_2$ with the Nakanishi-Lautrup
auxiliary field $B$. In other words, we have the freedom to choose
$B_2 \equiv B = - (\partial^\mu A_\mu)$. This ultimately leads to
\begin{equation}
\bar F^{(c)} (x, \bar\theta) = \bar C (x) + \bar\theta \;(\tilde
s_b \bar C (x)),\label{3.7}
\end{equation}
which is similar to the expansions of the other chiral superfields
in (\ref{3.6}).

We wrap up this subsection with the comment that the anti-chiral
version of the above discussion does not lead to any appropriate
nilpotent symmetry transformations (see, e.g. [20]). This
provides, in a subtle way, the reason behind the non-existence of
the on-shell nilpotent anti-BRST transformations for above cited
Lagrangian densities of the non-Abelian gauge theory. In contrast
to this observation, for the Abelian $U(1)$ gauge theory, it has
been shown [20] that the anti-chiral version of the above
superfield formalism does lead to the derivation of the on-shell
nilpotent (and anticommuting) anti-BRST symmetry transformations.

\subsection{Gauge-Invariant Condition: Nilpotent BRST
Symmetry Transformations for Matter Fields}

To obtain the nilpotent BRST transformations for the matter fields
$ \psi(x), \bar{\psi(x)}$, we begin with the following gauge
invariant condition on the chiral five (4,1)-dimensional super
sub-manifold of the (4, 2)-dimensional supermanifold:
\begin{equation}
\bar{\Psi}^{(c)} (x, \bar{\theta})\; (\tilde{d}_{(c)} + i \tilde
A_{(c)}^{(1)(h)})\; \Psi^{(c)} (x, \bar\theta) = \bar{\psi} (x)\;
(d + i A^{(1)}) \; \psi(x), \label{3.8}
\end{equation} where the
expansions for the chiral Dirac superfields are
\begin{equation}
\Psi^{(c)} (x, \bar\theta) = \psi(x) + i\; \bar{\theta}\;
(b_1\cdot T) (x), \qquad \bar{\Psi}^{(c)} (x, \bar\theta) =
\bar{\psi}(x) + i\; \bar{\theta}\; (b_2\cdot T) (x). \label{3.9}
\end{equation}
It is evident that the r.h.s of (\ref{3.8}) [i.e. $ dx^\mu\;
\bar{\psi}(x)\;(\partial_\mu +i A_\mu)\; \psi(x)$] is a gauge
(i.e. BRST)  invariant quantity for the non-Abelian gauge theory
described by the Lagrangian densities (\ref{2.1}), (\ref{2.6a})
and (\ref{2.6b}). On the l.h.s. of (\ref{3.8}), we have $\tilde
A^{(1)(h)}_{(c)} = dx^\mu ( A_\mu + \bar\theta D_\mu C) + d
\bar\theta [C - \frac{i}{2} \bar\theta (C \times C)]$ which is the
expression for the super 1-form $\tilde A^{(1)}_{(c)}$ after the
application of the horizontality condition.

It is straightforward to note that the l.h.s. of (\ref{3.8}) would
produce the coefficients of the differentials $dx^\mu, dx^\mu
(\bar\theta), d \bar\theta$ and $d \bar\theta (\bar\theta)$. Out
of which, the coefficient of $dx^\mu$ will match with a similar
kind of term emerging from the r.h.s. It is but natural that the
rest of the coefficients would be set equal to zero. For algebraic
convenience, it is useful to first collect the coefficients of
$d\bar\theta$ and $d \bar \theta (\bar\theta)$ from the l.h.s. of
(\ref{3.8}). These coefficients are
\begin{equation}
- i\; d\bar\theta\; [ \bar\psi b_1 + \bar \psi C \psi ] +
d\bar\theta \;(\bar\theta)\; [ b_2 b_1 + \frac{1}{2} \bar \psi (C
\times C) \psi + \bar \psi C b_1 + b_2 C \psi ]. \label{3.10}
\end{equation}
Setting equal to zero the coefficients of $d \bar\theta$ and $d
\bar\theta (\bar\theta)$ separately and independently, we obtain
(for $\bar\psi \neq 0$), the following relationships
\begin{equation}
b_1 \equiv b_1 \cdot T = - (C \cdot T) \psi, \qquad C b_1 +
\frac{1}{2} (C \times C) \psi = 0, \label{3.11}
\end{equation}
where, in the proof of the latter condition, the former condition
$b_1 = - (C \cdot T) \psi$ has been used as an input.

The explicit expressions for the coefficients of $dx^\mu$ and
$dx^\mu (\bar\theta)$, from the l.h.s. of (\ref{3.8}), are
collected together, as follows
\begin{equation}
dx^\mu\; (\bar{\psi}D_\mu\psi) -i\; dx^\mu \;(\bar{\theta})\; [
\bar{\psi} D_\mu  b_1 - b_2 D_\mu \psi + \bar{\psi}D_\mu C \psi ].
\label{3.12}
\end{equation}
It is clear that the coefficient of $dx^\mu$ matches with that of
the r.h.s. Using $b_1=-(C\cdot T)\psi$, we obtain the following
relationship
\begin{equation}
i \;dx^\mu\; \bar{\theta}\;[ \;\bar{\psi} C\cdot T +  b_2 \;]\;
D_\mu \psi = 0. \label{3.13}
\end{equation}
For our present interacting non-Abelian gauge theory, $D_\mu
\psi\neq 0$. Thus, the solution that emerges from (\ref{3.13}), is
\begin{equation}
b_2\cdot T = -\; \bar{\psi}\;(C\cdot T), \label{3.14}
\end{equation}
Substitution of $b_1$ and $b_2$ from (\ref{3.11}) and (\ref{3.13})
into (\ref{3.9}), leads to the derivation of the nilpotent BRST
symmetry transformations (\ref{2.3}) for the matter fields, as
given below
\begin{equation}
\Psi(x,\bar{\theta}) = \psi(x) + \bar{\theta}\;(\tilde{s}_b\psi),
\qquad \bar{\Psi}(x,\bar{\theta}) = \bar{\psi}(x) + \bar{\theta}
\; (\tilde{s}_b\bar{\psi}).\label{3.15}
\end{equation}
The above expansion, in terms of $\tilde s_b$, is exactly in the
same form as the expansions (\ref{3.6}) and (\ref{3.7}). It should
be emphasized that the gauge-covariant version of (\ref{3.8})
(i.e. $(\tilde d + i \tilde A^{(1)(h)}) \Psi^{(c)} = (d + i
A^{(1)}) \psi$) does {\it not} lead to the derivations of
(\ref{3.15}). Rather, it leads to a {\it unphysical} restriction:
$D_\mu \psi = 0$.

The expansions in (\ref{3.6}), (\ref{3.7}) and (\ref{3.15})
provide the geometrical interpretations for the on-shell nilpotent
BRST symmetry $\tilde{s}_b$ ( and the corresponding generator BRST
charge $\tilde{Q}_b$)as the translation generator
$(\partial/\partial\bar{\theta})$ along the Grassmannian direction
$(\bar{\theta})$ of the chiral five (4,1) dimensional super sub-
manifold (parameterized by $x^\mu $ and $ \bar{\theta}$) of the
general six (4, 2)-dimensional supermanifold. We note, in passing,
that there is a mutual consistency and complementarity between the
horizontality condition and the gauge invariant relation as (i)
they are inter-linked, and (ii) they owe their origin to the super
exterior derivatives $(\tilde{d}) d$ and super 1-form connection
$(\tilde{A}^{(1)}) A^{(1)}$.

\section{Off-Shell Nilpotent Symmetries: Augmented Superfield Formalism}

In this section, we shall derive the off-shell nilpotent symmetry
transformations for {\it all} the fields of the (anti-) BRST
Lagrangian densities given in (\ref{2.1}), (\ref{2.6a}),
(\ref{2.6b}) by applying the augmented superfield formalism.

\subsection{Horizontality Condition: Off-Shell Nilpotent BRST
and Anti-BRST Symmetries}

For the paper to be self-contained, we recapitulate very briefly
some of the key points of the earlier work [3] on the
horizontality condition in the context of non-Abelian gauge
theory. We consider the superfields $B_\mu (x, \theta,
\bar\theta), F (x, \theta, \bar\theta)$ and $ \bar F (x, \theta,
\bar\theta)$ that form the vector multiplet of the super 1-form
connection $\tilde A^{(1)} = d Z^M A_M = dx^\mu B_\mu + d\theta
\bar F + d \bar \theta F$ on the general six $(4, 2)$-dimensional
supermanifold. These component superfields can be expanded, in
terms of the basic fields $A_\mu, C, \bar C$ and the secondary
fields, along the Grassmannian directions of this general
supermanifold, as [3,4,11,20]
\begin{eqnarray}
(B_\mu \cdot T) (x, \theta, \bar\theta) &=& (A_\mu \cdot T) (x) +
\theta \bar (R_\mu \cdot T) (x) \nonumber\\&+& \bar\theta (R_\mu
\cdot T) (x) + i \theta \bar\theta (S_\mu \cdot T) (x),
\nonumber\\  (F \cdot T) (x, \theta, \bar\theta ) &=& (C \cdot T)
(x) + i \theta (\bar B_1 \cdot T) (x) \nonumber\\ &+& i \bar
\theta (B_1 \cdot T) (x) + i \theta \bar\theta (s \cdot T) (x),
\nonumber\\  (\bar F \cdot T) (x, \theta, \bar\theta) &=& (\bar C
\cdot T) (x) + i (\bar B_2 \cdot T) (x) \nonumber\\ &+& i \bar
\theta (B_2 \cdot T) (x) + i \theta \bar\theta (\bar s \cdot T)
(x). \label{4.1}
\end{eqnarray}
It will be noted that the bosonic fields $A_\mu, S_\mu, B_1, \bar
B_1, B_2, \bar B_2$ do match with the fermionic fields $R_\mu,
\bar R_\mu, C, \bar C, s, \bar s$ \footnote{Hereafter, we shall be
using, more often, the simpler notations $A_\mu \equiv A_\mu \cdot
T, B_2 \equiv B_2 \cdot T$, etc., for the sake of brevity in the
rest of the text.}. All the secondary fields will be expressed in
terms of the basic and auxiliary fields of the above cited 4D
(anti-) BRST invariant Lagrangian densities (\ref{2.6a}) and
(\ref{2.6b}) for the non-Abelian gauge theory by tapping the
potential of the horizontality condition on the general six $(4,
2)$-dimensional supermanifold.

For the application of the horizontality condition in its full
blaze of glory, it is important to define the super exterior
derivative $\tilde d = dx^\mu \partial_\mu + d \theta
\partial_\theta + d \bar \theta \partial_{\bar\theta}$ (with
$\tilde d^2 = 0$) on the six (4, 2) dimensional supermanifold.
Exploiting $\tilde d$ and $\tilde A^{(1)}$, one can define the
super 2-form curvature $\tilde F^{(2)} = \tilde d \tilde A^{(1)} +
i \tilde A^{(1)} \wedge \tilde A^{(1)}$ by exploiting the
Maurer-Cartan equation. This is subsequently equated with the
ordinary 2-form curvature $F^{(2)} = d A^{(1)} + i A^{(1)} \wedge
A^{(1)}$. This equality (the so-called horizontality condition)
yields the following relationships between the secondary fields on
one hand and the basic fields and auxiliary fields on the other
hand of the super expansion (\ref{4.1}) (see, e.g., [3,11,20] for
details)
\begin{eqnarray}
&& R_\mu = D_\mu C, \quad \bar R_\mu = D_\mu \bar C, \quad \bar
B_1 + B_2 = - (C \times \bar C), \quad \bar s = - i (B_2 \times
\bar C), \nonumber\\ && B_1 = -\; \frac{1}{2}\; (C \times C),
\qquad \bar B_2 = -\; \frac{1}{2}\; (\bar C \times \bar C), \qquad
s = i \;(\bar B_1 \times C), \nonumber\\ && S_\mu = D_\mu B_2 +
D_\mu C \times \bar C\; \equiv\; - D_\mu \bar B_1 - D_\mu \bar C
\times C. \label{4.2}
\end{eqnarray}
As we have done earlier in Sec. 3, we identify $B_2 = B$ and $\bar
B_1 = \bar B$ of the (anti-) BRST invariant Lagrangian densities
(\ref{2.6a}) and (\ref{2.6b}) of Sec. 2. Having done this, we
immediately obtain the (anti-) BRST transformations (\ref{2.2})
and (\ref{2.5}) because the super expansion in (\ref{4.1}) can now
be written, in terms of these transformations, as (see, e.g.
[3,11,20] for details)
\begin{eqnarray}
&& B^{(h)}_\mu (x, \theta, \bar\theta) = A_\mu (x) + \theta
(s_{ab} A_\mu (x)) + \bar \theta (s_b A_\mu (x)) + \theta
\bar\theta (s_b s_{ab} A_\mu (x)), \nonumber\\ && F^{(h)} (x,
\theta, \bar\theta) = C (x) + \theta \;(s_{ab} C (x)) +
\bar\theta\; (s_{b} C (x)) + \theta\; \bar\theta\; (s_b s_{ab} C
(x)), \nonumber\\ && \bar F^{(h)} (x, \theta, \bar\theta) = \bar C
(x) + \theta \;(s_{ab} \bar C (x)) + \bar\theta \;(s_b \bar C (x))
+ \theta\; \bar \theta \;(s_b s_{ab} \bar C (x)), \label{4.3}
\end{eqnarray}
where the superscript ${(h)}$, on the above superfields, denotes
the ensuing form of the superfields after the application of the
horizontality condition. The above equation, {\it vis-{\`a}-vis}
equation (\ref{2.7}), provides the geometrical interpretation of
the (anti-) BRST charges (that generate the off-shell nilpotent
($s_{(a)b}^2 = 0$) transformations $s_{(a)b}$) as the
translational generators along the Grassmannian directions of the
six (4, 2)-dimensional supermanifold.

\subsection{Gauge Invariant Condition: Off-Shell Nilpotent (Anti-)
BRST Symmetries for Matter Fields}

To obtain the off-shell nilpotent BRST and anti-BRST symmetries
{\it together} for the matter fields, we begin with the following
gauge (i.e. BRST) invariant condition on the six $(4,
2)$-dimensional supermanifold
\begin{equation}
\bar \Psi (x, \theta, \bar\theta) \bigl (\tilde d + i \tilde
A^{(1)(h)} \bigr ) \Psi (x, \theta, \bar\theta) = \bar \psi (x)
\bigl (d + i A^{(1)}) \psi (x), \label{4.4}
\end{equation}
where the superfields $\Psi (x, \theta, \bar \theta)$ and $\bar
\Psi (x, \theta, \bar\theta)$, corresponding to the 4D spinors
$\psi (x)$ and $\bar\psi (x)$, have the following expansions along
the Grassmannian directions of the six (4, 2)-dimensional
supermanifold
\begin{eqnarray}
\Psi (x, \theta, \bar\theta) &=& \psi (x) + i \theta (\bar b_1
\cdot T) (x) + i \bar\theta (b_1 \cdot T) (x)  + i \theta
\bar\theta (f \cdot T) (x), \nonumber\\ \bar \Psi (x, \theta,
\bar\theta) &=& \bar \psi (x) + i \theta (\bar b_2 \cdot T) (x) +
i \bar\theta (b_2 \cdot T) (x) + i \theta \bar\theta (\bar f \cdot
T) (x), \label{4.5}
\end{eqnarray}
where (i) bosonic (commuting) fields $b_1, \bar b_1, b_2, \bar
b_2$ do match with the fermionic (anticommuting) fields $\psi,
\bar \psi, f, \bar f$, (ii) the secondary fields $b_1, \bar b_1,
b_2, \bar b_2, f, \bar f$ will be determined, in terms of the
basic and auxiliary fields of the Lagrangian densities
(\ref{2.6a}) and (\ref{2.6b}), due to the gauge invariant
restriction (\ref{4.4}) invoked on the above general
supermanifold, and (iii) the explicit form of the super 1-form
connection $\tilde A^{(1)(h)}$, on the l.h.s. of equation
(\ref{4.4}), is: $\tilde A^{(1)(h)} = dx^\mu B_\mu^{(h)} + d
\theta \bar F^{(h)} + d \bar\theta F^{(h)}$. The expressions for
the redefined multiplet superfields $B_\mu^{(h)}, F^{(h)}, \bar
F^{(h)}$ are given in (\ref{4.3}).

It is clear that, on the r.h.s. of (\ref{4.4}), we have the gauge
(i.e. BRST) invariant quantity $(dx^\mu) [\psi (x) (\partial_\mu +
i A_\mu \cdot T) \psi (x)]$ with a single differential $dx^\mu$ on
the 4D spacetime sub-manifold of the six (4, 2)-dimensional
supermanifold. The l.h.s. will, however, produce the coefficients
of the differentials $dx^\mu, d \theta$ and $d \bar\theta$. Out of
these, only the coefficient of the pure $dx^\mu$ would match with
the r.h.s. The rest of the coefficients of all the differentials
will be set equal to zero. For algebraic convenience, it is
helpful and handy to collect the coefficients of $d \theta$ and $d
\bar\theta$ in the first go. The coefficients of $d\theta$ are
\begin{equation}
-\;i\; d \theta \;[\; \bar \psi (\bar b_1 + \bar C \psi)\; ] + d
\theta\; (\theta)\; L_1 + d \theta \;(\bar\theta)\; L_2 + d
\theta\; (\theta\bar\theta)\; L_3, \label{4.6}
\end{equation}
where the explicit expressions for $L_1, L_2$ and $L_3$ are
\begin{eqnarray}
L_1 &=& \bar b_2 \bar b_1\; + \bar b_2 \bar C \psi + \;\bar \psi
\bar C \bar b_1 +\; \frac{1}{2}\; \bar\psi (\bar C \times \bar C)
\psi, \nonumber\\ L_2 &=& b_2 \bar b_1 + i \bar \psi f + b_2 \bar
C \psi - \bar\psi B \psi + \bar\psi \bar C b_1, \nonumber\\ L_3
&=& \bar f \bar b_1 + \bar b_2 f + \bar \psi \bar C f + \bar f
\bar C \psi - i \bar \psi (B \times \bar C) \psi - i \bar \psi B
\bar b_1 \nonumber\\ &-& \frac{i}{2} \bar\psi (\bar C \times \bar
C) b_1 + i \bar b_2 B \psi - i \bar b_2 \bar C b_1 + i b_2 \bar C
\bar b_1 + \frac{i}{2} b_2 (\bar C \times \bar C) \psi.
\label{4.7}
\end{eqnarray}
Setting equal to zero the coefficients of $d \theta, d \theta
(\theta), d \theta (\bar\theta)$ and $d\theta (\theta\bar\theta)$,
we obtain the following solutions (for $\bar \psi \neq 0$)
\begin{equation}
\bar b_1 = - \;(\bar C \cdot T)\; \psi, \qquad f = - \;i \;(B \psi
- \bar C b_1). \label{4.8}
\end{equation}
The rest of the conditions, that emerge after the imposition of
the above restrictions, are satisfied if we use the above values
of $ \bar b_1$ and $f$.

In exactly similar fashion, collecting the coefficients of
$d\bar\theta$ from the l.h.s. of (\ref{4.4}), we get the following
explicit expressions
\begin{equation}
- \;i\; d \bar\theta\; [ \bar \psi (b_1 + C \psi) ] + d \bar\theta
\;(\theta)\; M_1 + d \bar\theta\; (\bar\theta)\; M_2 + d
\bar\theta\; (\theta \bar\theta)\; M_3, \label {4.9}
\end{equation}
where the explicit forms of $M_1, M_2 $ and $M_3$ are
\begin{eqnarray}
M_1 &=& \bar b_2 \;(b_1 + C \psi)  + \bar \psi \;(C \bar b_1 - i f
- \bar B \psi), \nonumber\\ M_2 &=& b_2\; (b_1 + C \psi) +
\bar\psi\; [ C b_1 + \frac{1}{2} (C \times C) \psi ], \nonumber\\
M_3 &=& b_2 [ f + i C \bar b_1 - i \bar B \psi ] + \bar f\; [ b_1
+ C \psi ] - i \bar b_2 [ C b_1 + \frac{1}{2} (C \times C) \psi ],
\nonumber\\ &+& \bar \psi\; \bigl [ C f + i (\bar B \times C) \psi
+ \frac{i}{2} (C \times C) \bar b_1 + i \bar B b_1 \bigr ].
\label{4.10}
\end{eqnarray}
Setting equal to zero the coefficients of $d\bar\theta$, $d \bar
\theta (\theta)$, $d \bar\theta (\bar\theta)$ and $d \bar\theta
(\theta\bar\theta)$ separately and independently, we obtain the
following solutions (for $\bar\psi \neq 0$)
\begin{equation}
b_1 = - (C \cdot T)\; \psi, \qquad f = i\; (\bar B \psi - C \bar
b_1). \label {4.11}
\end{equation}
It will be noted that the expressions for $f$, derived in
(\ref{4.8}) and (\ref{4.11}), are consistent because, finally,
they lead to $B + \bar B = - (C \times \bar C)$ that has already
been quoted in (\ref{4.2}). The rest of the relations are found to
be consistent if we use the values of $b_1, \bar b_1$ and $f$
given in (\ref{4.8}) and (\ref{4.11}).

Finally, we collect the coefficients of $dx^\mu$ from the l.h.s.
of equation (\ref{4.4}). These are written, in their most explicit
form, as follows
\begin{eqnarray}
&& dx^\mu\; ( \bar\psi \partial_\mu \psi + i \bar\psi A_\mu \psi)
\nonumber\\ && + i\; dx^\mu \;(\theta)\; [ \bar b_2 D_\mu \psi -
\bar\psi D_\mu \bar b_1 - \bar\psi (D_\mu \bar C) \psi ]
\nonumber\\ && + i \;dx^\mu (\bar\theta)\; [ b_2 D_\mu \psi -
\bar\psi D_\mu b_1 - \bar\psi (D_\mu C) \psi ], \nonumber\\ && +
i\; dx^\mu \;(\theta\bar\theta) \;\Bigl [ \bar f D_\mu \psi  + i
\bar b_2 ( D_\mu b_1 + D_\mu C \psi ) \nonumber\\ && - i b_2 (
D_\mu \bar b_1 + D_\mu \bar C \psi ) + \bar\psi \{ D_\mu f - i
D_\mu \bar C b_1 \nonumber\\ && + i D_\mu C \bar b_1 + i D_\mu B
\psi + i (D_\mu C \times \bar C) \psi \} \Bigr ]. \label{4.12}
\end{eqnarray}
It is obvious that the coefficient of  pure $d x^\mu$ does match
with the corresponding coefficient emerging from the r.h.s.
Exploiting the values of $\bar b_1, b_1$ and $f$ from (\ref{4.8})
and (\ref{4.11}), and setting the coefficients of $dx^\mu
(\theta), dx^\mu (\bar\theta)$ and $dx^\mu (\theta\bar\theta)$
equal to zero separately, do lead to the following conditions
\begin{eqnarray}
&&(\bar b_2 + \bar \psi \bar C)\; D_\mu \psi = 0, \qquad (b_2 +
\bar\psi C)\; D_\mu \psi = 0, \nonumber\\ && \bigl [\bar f - i
\bar b_2 C + i b_2 \bar C - i \bar \psi \{B + \frac{1}{2} (C
\times \bar C) \} \bigr ] \; D_\mu \psi = 0. \label{4.13}
\end{eqnarray}
It is clear that, for the interacting non-Abelian gauge theory
under consideration, we have $D_\mu \psi \neq 0$ because it leads
to the existence of interaction term. Thus, the solutions we
obtain from the above,  are
\begin{equation}
\bar b_2 = - \bar \psi \bar C, \quad b_2 = - \bar \psi C, \quad
\bar f = i\; \bar\psi\; \bigl [ B + \frac{1}{2} (C \times C) \bigr
]. \label{4.14}
\end{equation}
Substitutions of the values from (\ref{4.8}), (\ref{4.11}) and
(\ref{4.14}) into the super expansions (\ref{4.5}), lead to the
following
\begin{eqnarray}
\Psi (x, \theta, \bar\theta) &=& \psi (x) + \theta (s_{ab} \psi
(x)) + \bar \theta (s_b \psi (x)) + \theta \bar\theta (s_b s_{ab}
\psi (x)), \nonumber\\ \bar \Psi (x, \theta, \bar\theta) &=& \bar
\psi (x) + \theta (s_{ab} \bar \psi (x)) + \bar\theta (s_b \bar
\psi (x)) + \theta \bar\theta (s_b s_{ab} \bar\psi (x)),\label
{4.15}
\end{eqnarray}
where the (anti-) BRST transformations $s_{(a)b}$ are illustrated
in (\ref{2.2}) and (5). Thus, we obtain the above symmetry
transformations for the matter fields {\it together}. The
geometrical interpretations for the (anti-) BRST charges and the
corresponding transformations $s_{(a)b}$ are same as the ones
given for the gauge and (anti-) ghost fields in Sec. 3.

\section{Conclusions}

The derivation of the mathematically exact expressions for the
nilpotent and anticommuting (anti-) BRST symmetry transformations,
associated with the matter fields of an interacting (non-) Abelian
gauge theory, has been an outstanding problem in the framework of
superfield approach to BRST formalism. In our present endeavour,
we have been able to resolve this long-standing problem because we
have been able to derive the exact forms of the nilpotent (anti-)
BRST symmetry transformations for the matter (Dirac) fields of a
$SU(N)$ non-Abelian gauge theory where there is an interaction
between the non-Abelian gauge field and the matter (Dirac)
fields\footnote{ In fact, it is the conserved matter Noether
current that couples to the gauge fields of the 1-form (non-)
Abelian gauge theories to generate the interaction term when one
requires the local gauge invariance in the theory (see, e.g. [21]
for details).}. In fact, the physical insights into the gauge
(i.e. BRST) invariant quantities (cf. (\ref{3.8}) and (\ref{4.4}))
have helped us to obtain the proper restrictions on the five (4,
1)-dimensional chiral super sub-manifold and the general six
$(4,2)$-dimensional supermanifold which lead to the above exact
derivations.

The above cited gauge (i.e. BRST) invariant quantities originate
from the key properties associated with the (super) covariant
derivatives on the supermanifold. Some of the  striking
similarities and key differences between the horizontality
condition and the gauge invariant condition(s) are as follows.
First, both of them primarily owe their origin to the (super)
cohomological operators $\tilde d $ and $d$. Second, the
geometrical origin and interpretations for the (anti-) BRST
charges (and the nilpotent symmetry transformations they generate)
remain intact for the validity of both the conditions on the
supermanifold. Third, whereas the horizontality condition is an
$SU(N)$ covariant restriction (because $F^{(2)} \to U F^{(2)}
U^{-1}$ where $U \in SU(N)$), the other condition(s), as the name
suggests, are the $SU(N)$ gauge invariant condition(s). Finally,
as mentioned in the Sec. 3.2, the covariant versions of the
gauge-invariant restrictions (cf. (\ref{3.8}) and (\ref{4.4})) do
not lead to the exact derivations of the nilpotent symmetry
transformations whereas the horizontality condition (basically a
covariant restriction) does lead to exact derivations of the
nilpotent symmetry transformations for the gauge and (anti-)ghost
fields of the Lagrangian density of a non-Abelian gauge theory.

In our earlier works [8-13], we have consistency extended the
horizontality condition by requiring the equality of the
supersymmetric versions of the conserved currents/charges with the
ordinary local conserved corrents/charges. In one of our recent
works [13], in addition to the horizontality condition, any
conserved quantities are required to be invariant on the
supermanifold. In the present work and some of its precursors
[14,15], we have exploited the (super) covariant derivatives in
the construction of the gauge (i.e. BRST) invariant quantities
which have been equated on the certain specific supermanifolds.
The latter ones yield the exact expressions for the nilpotent
symmetries for the matter fields whereas the former ones lead to
the consistent derivations of the same. These consistent and
complementary extensions of the horizontality condition have been
christened as the {\it augmented} superfield approach to BRST
formulation because they yield transformations for {\it all} the
fields.

One of us, in his earlier works [22-26], has studied in detail,
the gauge theories in a different type of superspace (also called
BRST superspace). The salient features of these works are (i) the
whole action including the source terms for the composite
operators is accommodated in a single compact superspace action,
(ii) the theory has a generalized gauge invariance in superspace,
(iii) the superspace is completely unrestricted  such that the
operations like super rotation and translation, in anticommuting
coordinates, can be carried out, and (iv) the WT identities are
realized in a very simple manner [22-24]. These superspace
formulations are used to study renormalization of gauge theories,
in particular, the renormalization of gauge invariant operators
[23-27]. It would be an interesting endeavour to study similar
things in the present superfield formulation. Recently, the
results of the covariant horizontality condition for Abelian
$U(1)$ gauge theory have been derived from a gauge invariant
condition involving covariant derivatives, Dirac fields and their
connection with (super) 2-forms $(\tilde F^{(2)})F^{(2)}$ [28]. It
would be a nice endeavour to check the same for the non-Abelian
gauge theory. Furthermore, it will be very interesting to find out
a single restriction on the supermanifold that could produce the
results of the horizontality condition and the new gauge invariant
condition(s) {\it together}. These are some of the issues that are
under investigation and our results would be reported elsewhere
[29].\\

\end{document}